# Thermal IR satellite data application for earthquake research in China

ANDREW A. TRONIN

Earth Observation Research Centre, National Space Development Agency of Japan (NASDA)

Roppongi 1-9-9, Minato-ku, Tokyo 106-0032 JAPAN

e-mail: tronin@eorc.nasda.go.jp

Telephone: +81-3-3224-7086

Facsimile: +81-3-3224-7052

**Abstract**. NOAA/AVHRR satellite thermal images indicated the presence of positive thermal anomalies that are associated with the large linear structures and fault systems of the Earth's crust. The relation between thermal anomalies and seismic activity was established for Middle Asia on the basis of a 7-year series of thermal images. Thermal anomaly has been located near Beijing, on the border between mountains and plain. The size of this anomaly is about 700 km in length and 50 km in width. The anomaly appeared about 6 – 24 days before and continued about a week after an earthquake. The anomaly was sensitive to crust earthquakes with a magnitude more than 4.7 and for distance of up to 500 km. The amplitude of this anomaly was about 3° C.

1. Introduction

The first application of thermal images in seismology was in Russia and started in 1985 and first results was published in 1988 (Gorny *et al.* 1988). Analogue Advanced Very High Resolution Radiometer (AVHRR) data, transmitted from National Ocean and Atmospheric Administartion (NOAA) satellite in analog format were analysed. The area of study was the seismically active region of Central Asia: Tien-Shan, Kizilkum and Karakum deserts, South Kazakhstan, (Tronin 1996, 1999). Later similar studies were carried out in China (Qiang *et al.* 1999), Greece, Japan and Spain. To develop this research further a 'Satellite Prediction Research Centre for Natural Disasters' was established in China.

2. Data

The north-east China was selected as appropriate test site: high seismicity, flat relief. (figure 1).

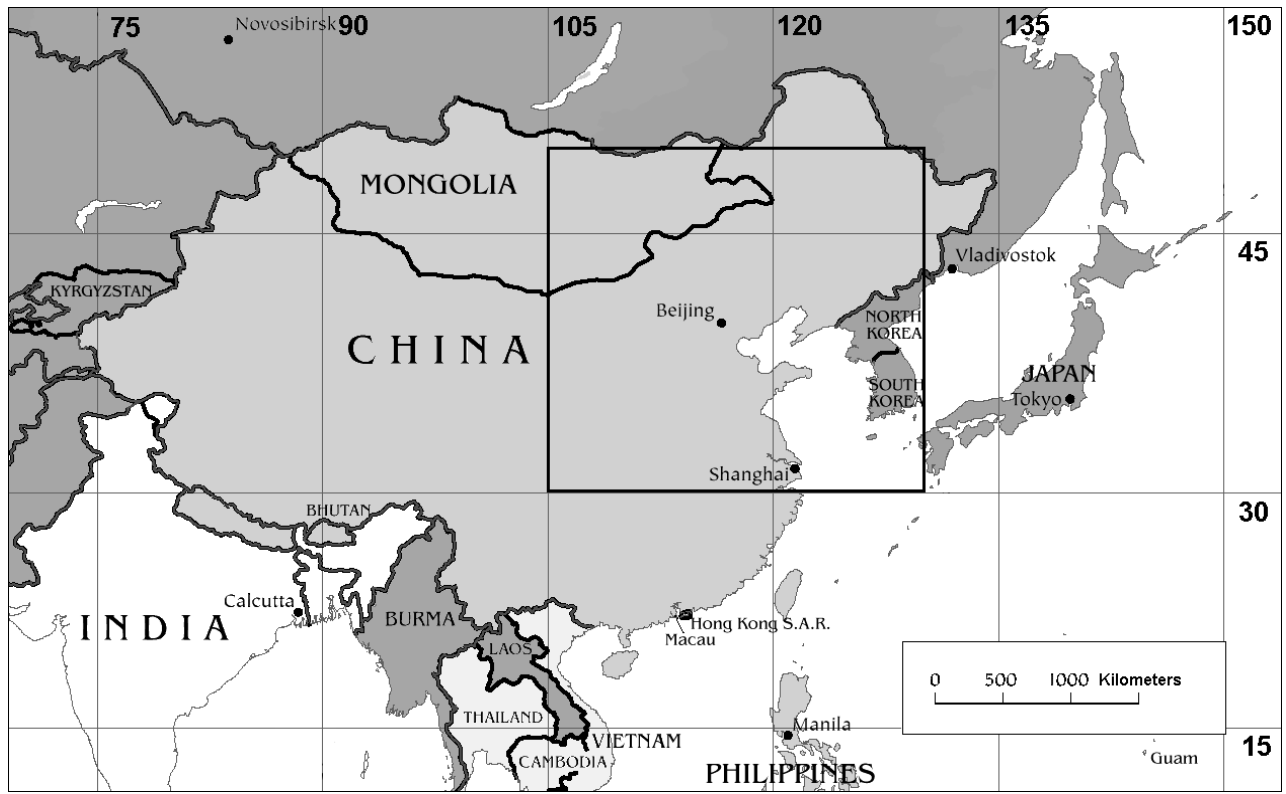

Figure 1. North-east China, area of research.

The period of observation: August 1998 – February 1999. NOAA satellite data, collected in Tokyo University (Kitsuregawa Laboratory) were used. For August – October 1998 – 41 digital images were processed. For the second period: December 1998 – February 1999 – 53 images. More then 200 quicklook images also were downloaded. The earthquake catalogue for north-east China is shown on table 1.

Table 1. The earthquake catalogue for north-east China. USGS catalogue, 30 – 50N, 105 – 130E, 1.08.98 – 1.03.99, M > 3.0.

| Date | Time | N | E | Depth, km | Magnitude | Area | Distance to thermal anomaly, km | Thermal anomaly response |
|---|---|---|---|---|---|---|---|---|
| 13.08.98 | 10:21:55.44 | 41° 9' | 114° 31' | 33 | 4.3 | China, Inner Mongolia | 200 | - |
| 29.08.98 | 20:14:04.08 | 43° 25' | 108° 59' | 33 | 4.7 | South Mongolia | 700 | - |
| 06.09.98 | 13:26:07.24 | 40° 45' | 120° 10' | 33 | 4.1 | China, Yellow Sea Area | 350 | - |
| 24.09.98 | 18:53:40.18 | 46° 19' | 106° 17' | 33 | 5.5 | Central Mongolia | 900 | + |
| 12.11.98 | 18:30:40.01 | 30° 8' | 129° 58' | 33 | 3.6 | Japan | 1700 | - |
| 18.12.98 | 10:08:34.29 | 30° 10' | 129° 58' | 123 | 4.7 | Japan | 1700 | - |
| 29.01.99 | 5:44:23.88 | 44° 40' | 115° 43' | 10 | 4.9 | China, Inner Mongolia | 400 | + |
| 02.02.99 | 13:22:21.3 | 49° 25' | 105° 33' | 33 | 3.7 | North Mongolia | 1300 | - |

3. Method

Image processing includes standard procedures for AVHRR thermal infrared (IR) images: data extraction, calibration, radiation and atmospheric corrections. The ground temperature was retrieved according Bekker's algorithm, using bands 4 and 5 (Becker and Li 1990). The emmissivity was set to 0.98 for both bands. As a result of first step temperature images of the Earth surface were obtained. On the next step thermal anomaly was selected, its area and temperature was measured. The background area on the plain, between Beijing and Huang He (Yellow River) was selected (figure 3). During the night a plain is the warmest place onshore. All pixels in similar meteorological conditions with a temperature exceeded average value could be considered as an anomaly. The average value and standard deviation for background area were calculated. Then all values on the image below average level plus two standard deviation were removed. Water surfaces and clouds were masked and excluded. Thus the thermal anomaly image was obtained. On this image the anomaly area and temperature were measured and registered to database. Mainly night data were used. If night data were absent, morning images were involved in processing.

4. Results

Preliminary results show the presents of IR anomaly in the similar geological and tectonic conditions to Middle Asia anomalies. We have founded the reaction on 5 earthquakes in north-east China (figure 2). Two examples of thermal anomalies are described below. One example of thermal anomalies in China is shown on figure 2.

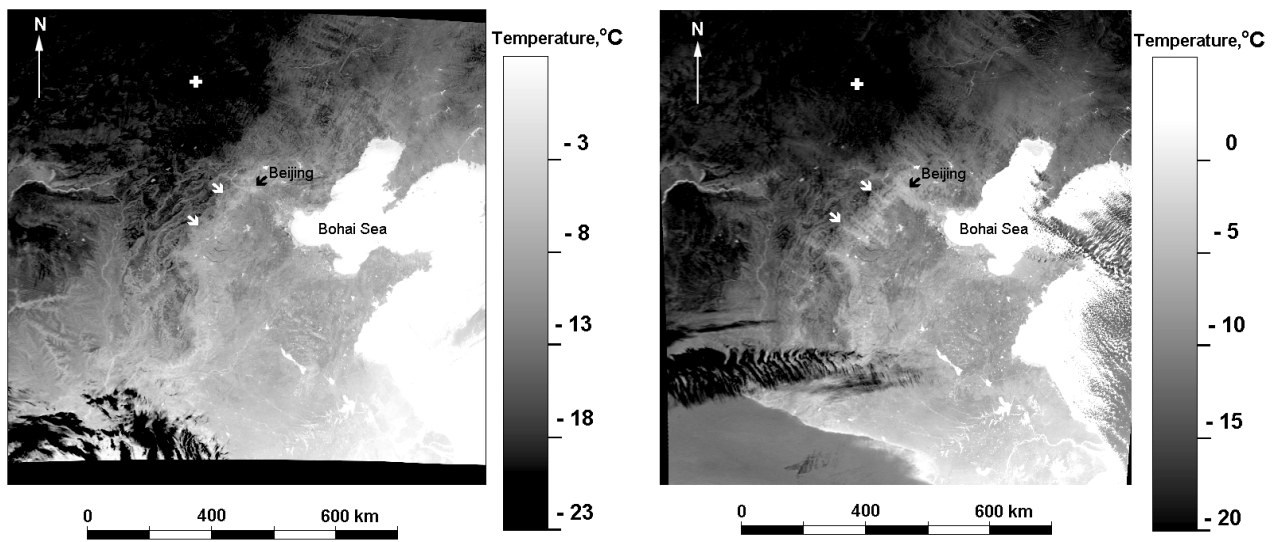

Figure 2. NOAA thermal images; (left) north-east China, background level, NOAA – 14, 28 Dec 1998, 19:21 GMT; (right) north-east China, thermal anomaly, NOAA – 14, 7 Jan 1999, 19:10 GMT. Arrows show thermal anomaly, cross – earthquake epicentre. Earthquake: 29 Jan 1999, 5:44 GMT, 44° 40' N 115° 43' E, h=10 km, M=4.9.

The anomaly located on the tectonic and geomorphologic border between mountain area (north parts of Shanxi and Hebei provinces, Wutai and Taihang mountains) and plain area (south part of Hebei province or Beijing plate) (figure 3). The anomaly is located in the geological and tectonic conditions similar to the Middle Asia anomalies (Tronin 1996, 1999). One can suppose the presents in this place large fault system, which could be the source of thermal anomaly.

Two sections were marked on the thermal image (figure 3). The section 1 shows the location and magnitude of the thermal anomaly. The left side of the profile lies in the Wutai mountains with altitudes 1500 – 2000 m and has low temperature: -15° – -20° C. Beijing plain – on the right side of the section – has the heights about 100 m and temperatures -10° C. The thermal anomaly locates between mountains and plain (with temperature up to -2° C), the dash line shows the background temperature trend. In background conditions (section 2) the temperature is fluently increases from mountains to plain according the altitude change.

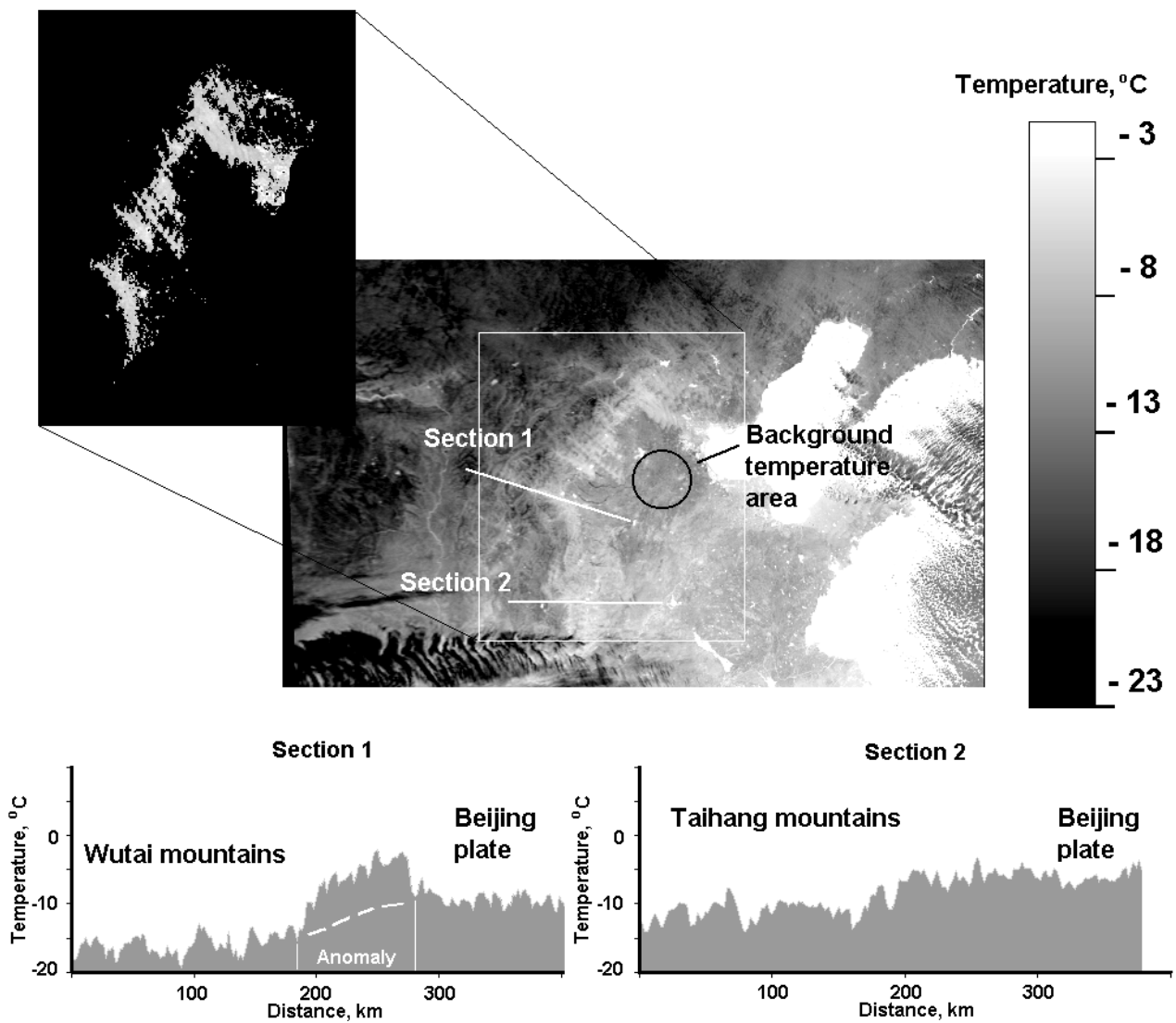

Figure 3. Thermal image, north-east China (top) and thermal sections (bottom). The image: NOAA –14, 7 Jan 1999, 19:10 GMT. Background area calculation: mean temperature is - 9.4° C, standard deviation is 1.2° C. The incut shows the thermal anomaly selection (values above the mean level plus two standard deviation). Section 1 – anomaly situation, Section 2 – background situation.

The described anomaly in background situation looks like the typical thermal anomaly caused by vertical distribution of air temperature at night (inverse temperature distribution). In stable night meteorological conditions we have relatively cold air near the Earth surface on the plain. The temperature increases with the altitude. The following height increase leads to temperature decrease. The meteorological anomaly strongly related to local relief. In our case the thermal anomaly spreads both the valley and heights, so it is impossible to explain the origin of anomaly by topographical reason.

As was described above the area with background temperatures was selected in Beijing plate. Then all necessary calculations were executed. The area and temperature of anomaly was set to database. The results for December 1998 – February 1999 are shown on figure 4.

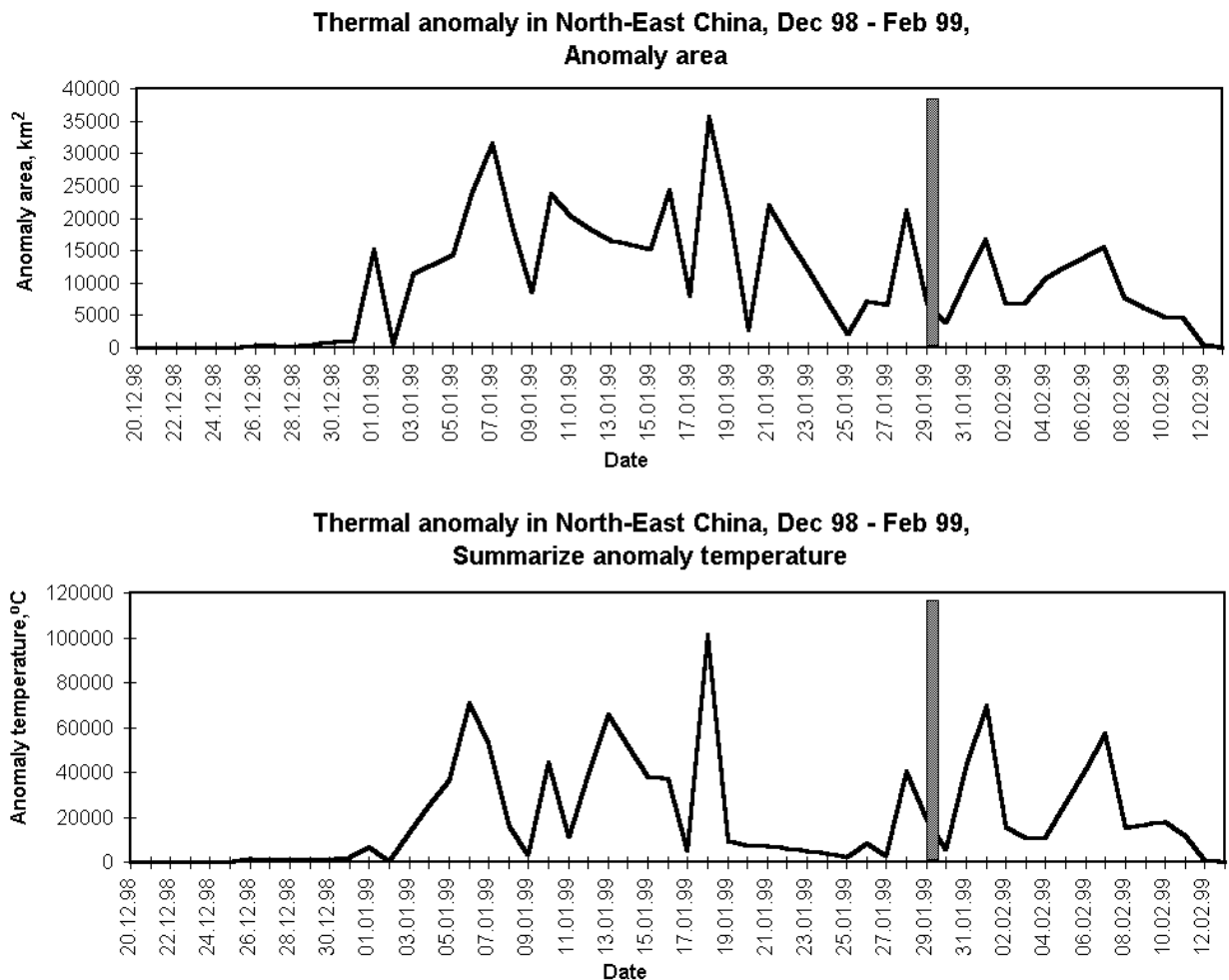

Figure 4. Thermal anomaly in north-east China related with earthquake. December 1998 – February 1999. Earthquake: 29 Jan 1999, 5:44 GMT, 44° 40' N 115° 43' E, h=10 km, M=4.9.

As one can see the anomaly appears clearly in 5 – 7 Jan 1999, about 3 week before shock. A few days before earthquake the size and temperature of anomaly were reduced. After the shock the temperature and area increased again. All these features are similar to thermal anomalies observed in Middle Asia (Tronin 1999).

The long series of observation (August 1998 – March 1999) confirms the correlation of thermal anomaly and seismic activity (figure 5).

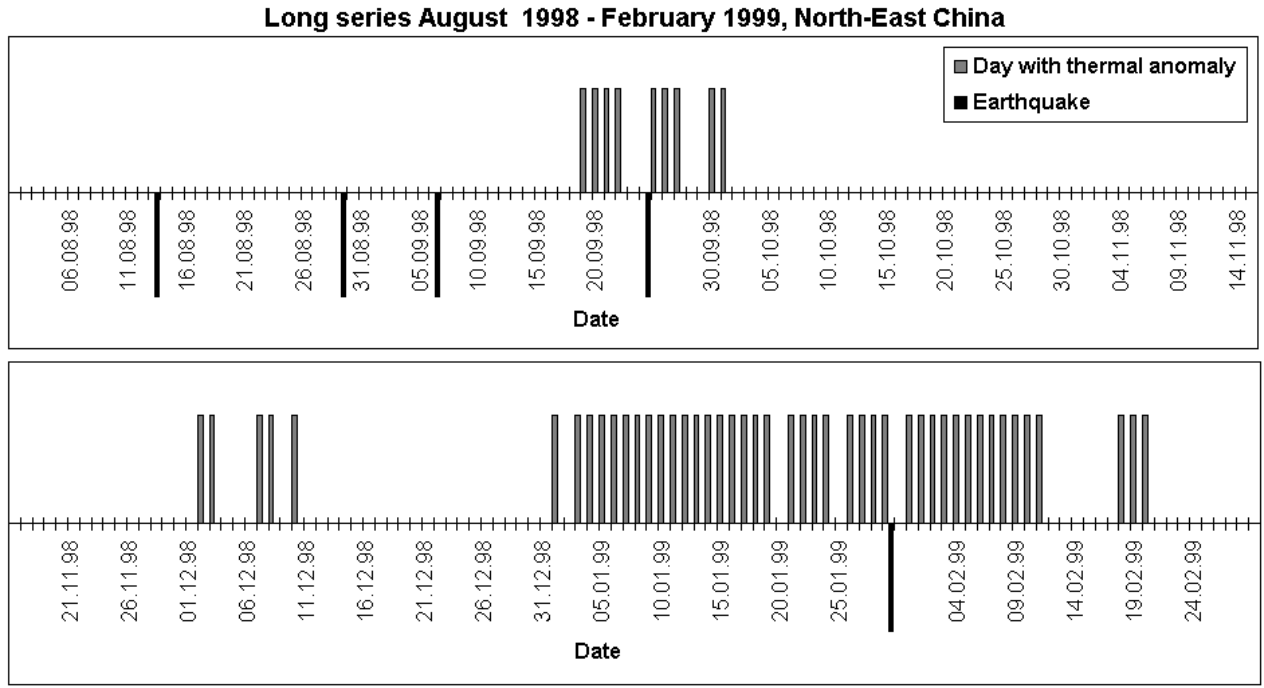

Figure 5. Thermal anomaly in north-east China. Long series of observation (August 1998 – March 1999). Earthquakes: 13.08.98, 29.08.98, 06.09.98, 24.09.98, 29.01.99 (see table 1).

As it is clear from the figure 5 there were no reaction of the thermal anomaly to three earthquakes: 13.08.98 M=4.3, China, Inner Mongolia; 29.08.98 M=4.7, south Mongolia; 06.09.98 M=4.1, China, Yellow Sea Area (table 1). Three other events in Japan and North Mongolia are remote earthquakes and do not affect to thermal anomaly. All these earthquakes have relatively low magnitude and locate too far from thermal anomaly area. Thermal anomaly related to the earthquake of 29.01.99 (M=4.9) looks to be more pronounced than the anomaly related to earthquake 24.09.98 (M=5.5). The explanation is the distance to thermal anomaly, in the first case it is 400 km, in second one – 900 km. This information will be use to determine the sensitivity of thermal anomaly to seismic events. The distance about 1000 km is probably the limit of thermal anomaly sensitivity in north-east China.

## 5. Discussion

At present the nature of IR anomalies is not clear. In all probably hydrogeological factor or greenhouse effect play main role in the forming of anomalies. The following equations of heat balance on the Earth surface could be written:

$$dT/dz \,|_{z=0} = q_r + q_{an} + q_d + q_t + q_{ev} + q_g \qquad (1)$$

where $q_r$ is the radiation balance (incoming and outgoing radiation); $q_{an}$ the heat flux in the soil induced by annual thermal rhythm; $q_d$ the heat flux in the soil induced by diurnal thermal rhythm; $q_t$ is the heat loss due to turbulent exchange between atmosphere and ground surface ( heat loss for wind ); $q_{ev}$ is the heat loss for evaporation; $q_g$ is geothermal flux, $z$ – depth.

Up to this moment the value of the geothermal flux $q_g$ was considered to be less than the other terms in equation (1) by several orders of magnitude. In this case the value of geothermal flux in the Central Asian region do not exceed $0.1 – 0.2$ $Wm^{-2}$. Therefore the variation of this parameter can not explain the origin of the surface temperature anomaly. However it is only the conductive component of geothermal flux. It is well known that at the Earth's surface the geothermal flux includes two components: the conductive and the convective. The convective flux is heat flux related to water and gas movement. Our current research shows that the convective component can be very high in some areas – $n*10$ $Wm^{-2}$. Some geological structures, e.g. rift-origin ones are characterised by abnormally high values of heat flow. Geothermal flux affects surface temperature as well as solar and meteorological heat fluxes.

The moisture content in soil and humidity in air remain very important factors controlling surface temperature. These parameters affect the run of such processes as an evaporation and moisture condensation – $q_{ev}$. The evaporation is most intensive in the daytime, when solar heating takes place, and it leads to a decrease of surface temperature. The moisture content in soil also alters its thermophysical properties and affects the process of dewfall, which is known to be associated with release of a heat.

Changes in the gaseous composition of the atmosphere can result in different effects. One of these is the greenhouse effect. Optically active gases such as $CO_2$, $CH_4$ and etc. absorb a part of the Earth's infrared radiation that leads to accumulation of the heat near the surface. Not so numerous observations of gas content were executed in open air (table 2). These measurements took place in open air with strong wind, a few hours after the shock in epicentre.

Therefore, anomalous concentration of optically active gases $CO_2$, $CH_4$, water vapour, in near-surface layers of the atmosphere could generate additional heat along anomaly zones and as a result could be considered as a possible source of the anomalous outgoing infrared radiation. Observed values of the temperature do not contradict values of temperature variations calculated according to the increase of gas concentration in the near-surface atmosphere. Kong *et al*. (1999) suggests promising hypothesis explaining thermal anomaly by the interaction of electric field and gas cloud.

Table 2. Air content in epicentre of Dagestan earthquake. 14 May 1970, 18:12 GMT, 43° N, 47° E, h=44 km, M=6.7. Percent by volume, (Osika 1981).

|  | $H_2$ | He | $O_2$ | $N_2$ | $CO_2$ | $CH_4$ |
|---|---|---|---|---|---|---|
| Fissure 1 | 0.038 | Up to 0.001 | 20.82 | 78.90 | Up to 0.1 | 0.00013 |
| Fissure 2 | 0.014 | Up to 0.001 | 20.31 | 79.58 | Up to 0.1 | 0.00014 |
| Background concentration | 0.00005 | 0.000524 | 20.6 | 78 | 0.035 | 0.00015 |

The hypotheses, discussed above, do not disagree with each other. By our opinion, mechanisms lead to the increase of soil moisture and gas concentration is very similar. Ground water, gas and heat motion (convective heat flux) may be unit reason for all effects.

6. Conclusion

Preliminary results of thermal anomaly investigations in China indicate: 1) the anomalies appeared about 6 – 24 days before and continued about a week after earthquake, 2) the anomaly was sensitive to crust earthquakes with a magnitude more than 4.7 and for distance of up to 500 km, 3) the size of anomaly is about 700 km in length and 50 km in width, 4) the amplitude of the anomaly is about 3° C.

Acknowledgement

The current research, presenting in this paper is executing in the framework of the project: 'Earthquake Remote Sensing Frontier Research Project' (NASDA). I would like to express my thanks to professors M. Hayakawa and O. Molchanov for help in support of this research.